 \documentclass{aa}  
\usepackage{graphicx}
\usepackage{longtable}
\usepackage{lscape}
\usepackage{txfonts}
\begin{document}
\title{A catalog of planetary nebula candidates and HII regions in NGC 3109\thanks{Based on observations collected at the European Southern Observatory, VLT, Paranal, Chile, program ID 076.B-0166}}
\author{M. Pe\~na\inst{1}\thanks{On sabbatical leave  at the Departamento de Astronom{\'\i}a, Universidad de Chile.}
\and
M. G. Richer\inst{2}
\and
G. Stasi\'nska\inst{3}
}
\offprints{M. Pe\~na}
\institute{Instituto de Astronom{\'\i}a, Universidad Nacional Aut\'onoma de M\'exico, Apdo. Postal 70264, M\'ex. D.F., 04510 M\'exico\\
\email{miriam@astroscu.unam.mx}
\and
Instituto de Astronom{\'\i}a, Universidad Nacional Aut\'onoma de M\'exico, Apdo. Postal 877,  Ensenada, B.C., M\'exico\\
\email{richer@astrosen.unam.mx}
\and
LUTH, Observatoire de Meudon, Meudon Cedex, France\\
\email{grazyna.stasinska@obspm.fr}
}
\date{Received 8 November 2006 / Accepted 30 December 2006 }


\abstract
{}
{Images obtained with the ESO VLT and FORS1 in [O {\sc iii}] 5007 on- and off-band, as well as  r\_Gunn filters, are analyzed to search for planetary nebula (PN) candidates.}
{In  the continuum-subtracted [O {\sc iii}] 5007 on-band images, a large number of emission-line regions were detected.
We describe the criteria employed for distinguishing  PN candidates from compact  HII regions.}
{ The most unambiguous discriminators for the two classes of nebulae are the sizes and properties of the ionizing stars. Based upon these criteria, we have found 20 PN candidates for which we present coordinates, nebular [O {\sc iii}] fluxes, and  stellar magnitudes.  The cumulative luminosity function for these PNe is discussed. A catalog of HII regions listing coordinates, nebular [O {\sc iii}] fluxes, stellar magnitudes, and other characteristics is also presented. We find that HII regions are rather concentrated towards the disk of the galaxy, while PNe are distributed  also above and below this structure, consistent with their belonging to an older stellar population.
}
{}

\maketitle
\section{Introduction}

The dwarf galaxies of the Local Group (LG) and its vicinity allow detailed studies of their different component populations, such as stars, planetary nebulae (PNe), and HII regions. Therefore, they provide strong constraints for stellar population studies and help  in understanding how the elemental abundances of galaxies have changed over time. Dwarf irregulars, in particular, allow a comparative analysis of the chemical properties of their stellar, PN, and HII region populations, which  provides a probe of the chemical enrichment and star formation history of the galaxy
from intermediate ages to the present day. In addition, through spectroscopic analysis of PNe and HII regions in different zones of these galaxies, it is possible to address the question of the large-scale homogeneity in the distribution of heavy elements. To proceed with these studies, the first step is to detect PNe and HII regions, distinguishing them from each other, which is a non-trivial matter in star-forming galaxies, before obtaining spectrophotometry  of the candidates. 
Spectroscopic analyses have already been done for several galaxies of the Local Group. For example, Stasi\'nska et al.  (1998) present a comparative analysis of the PN populations in  the Magellanic Clouds, M32, and the bulges of the Milky Way and M31. M33 was analyzed by Magrini et al. (2003), while Sextans A and B  were studied by Magrini et al. (2005) and Kniazev et al. (2006).

Here and in a forthcoming paper, we propose  studying the PN and HII region populations in NGC~3109 (DDO 236),  a Magellanic-type irregular classified as a late-type spiral SB(s)m (de Vaucouleurs et al. 1991).  NGC 3109  is the dominant member of the small group of metal-poor dwarf galaxies that also includes Sextans A, Sextans B, and Antlia, just at the edge of the LG. Due to its distance, NGC 3109 is particularly well-suited to chemical evolution studies because  it does not seem to have  been affected by tidal effects from the massive galaxies of the LG, so it should not have lost a significant amount of gas.

With an absolute magnitude  M$_{\rm B} \sim -$15.2 (see \S 4) and a mean oxygen abundance log O/H+12 $\sim$ 7.75 (as derived from HII regions by Lee et al. 2003  and, recently, from blue supergiants by Evans et al. 2006), NGC~3109  is similar to the SMC in luminosity and chemical composition. The galaxy is seen almost edge-on, having an elongated shape  aligned almost E-W, with an extension of  about 17.4$\times$3.5 arcmin. Studies of the stellar populations in this galaxy have been made by, among others, Demers et al. (1985),   who pointed out some traces of spiral structure. Sandage \& Carlson (1988)  reported light curves and periods for 29 cepheids and estimated a distance modulus of 26.0 mag (1.59 Mpc). The cepheids were revisited by Capaccioli et al. (1992), who derived a distance modulus of 25.5$\pm$0.16 (1.26$\pm$0.10 Mpc), and by Musella et al. (1997), who find a distance modulus of 25.67$\pm$0.16 (1.36$\pm$0.10 Mpc).  Based upon V and I band photometry, Minniti et al. (1999) established the presence of a halo consisting of Population II stars and derived a distance modulus of 25.62$\pm$0.1 mag (1.33$\pm$0.10 Mpc) from a deep luminosity function of these stars.  We adopt this value for the distance to NGC 3109 in the following sections.

NGC~3109 presents all of the usual indicators of robust star formation activity:  numerous extended and compact HII regions,  large ionized shells, super-shells,  and filaments of low surface brightness.  A catalog of extended HII regions  was published by Hodge in 1969. The OB associations and the distribution of star-forming regions were analyzed by Bresolin et al.  (1993), who also produced a catalog of HII regions. Hunter et al. (1993) analyzed the shells and super-shells in this galaxy.  

Richer \& McCall (1992) reported the discovery of  seven PN candidates and ten  HII regions in the central zone of NGC~3109 and,  from the brightest-PNe method, they derived a distance modulus of 25.96 (+0.1/$-$0.4) corresponding to 1.6 Mpc. From good-quality imaging Pe\~na et al. (2006; and this paper) have found that most of the PN candidates found by Richer \& McCall are more likely compact HII regions that are somewhat brighter than PNe, which explains why their distance modulus is overestimated compared to the stellar studies.  Leisy et al. (2005) report the discovery of 18 PN candidates in NGC 3109 and, from spectroscopy, they find that PNe display a wide range of metallicities.

Here, we present a deep survey of PNe and HII regions in NGC~3109 based on [O {\sc iii}]~5007 on- and off-band images, as well as {\it r\_Gunn} imaging covering almost the entire galaxy. The data were acquired with the ESO Very Large Telescope (VLT), equipped with the Focal reducer spectrograph FORS1.  
These images constitute the `pre-imaging' observations for the spectroscopic program ID 076.B-0166.  They were obtained  to identify candidates for a subsequent multi-object spectroscopy. 
From the continuum-subtracted [O {\sc iii}]~5007 images, we identified  more than 40 HII regions and 20 PN candidates that are reported in this work.

 The VLT-FORS1 multi-object spectroscopic run associated with the pre-imaging  observations was performed on January 27,  2006. Spectra of several HII regions and PN candidates, distributed in different zones of the galaxy, were acquired.  These data allowed us to confirm the PN nature of many PN candidates and will form the basis of a forthcoming  paper (Pe\~na et al., in preparation). Nonetheless the calibrated [O {\sc iii}] 5007 fluxes are presented here, as they allow us to calibrate the [O {\sc iii}] on-band imaging and help to strengthen our arguments. In our next paper,  we  shall study the  chemical abundances of nebulae and star formation history of the galaxy, along with  the large-scale spatial distribution of the heavy elements.

 This paper is organized as follows: in \S 2 we present the  'pre-imaging' observations and discuss the criteria we use to distinguish between planetary nebulae and compact HII regions.  In \S3, we present the properties of  the detected PNe and HII regions, including positions, diameters, [O {\sc iii}]~5007 calibrated fluxes, nebular magnitudes, and V magnitudes for the central stars.  
The cumulative luminosity function for the PNe is calculated and discussed in \S4,  and the conclusions presented in \S 5.

\section{Imaging:  distinguishing between PNe and compact HII regions}

The wide field coverage of FORS1  allowed us to observe almost the entire galaxy with only two frames. The observing program consisted  of imaging with the filters [O {\sc iii}] 5007 on-band (filter name  FILT\_500\_5+85, $\lambda_c$=5000 \AA, FWHM=50 \AA), [O {\sc iii}]  off-band (filter name OIII/6000+52, $\lambda_c$= 5105 \AA, FWHM=61 \AA) and {\it r\_Gunn} (hereafter {\it r}). The images were acquired in service mode on  November 29, and  December 1,  2005.

Two fields were observed, centered at 10:03:19.8  -26:09:32 (east field; ob ID 204405) and 10:02:54.5 -26:09:22 (west field; ob ID 204406). Each field has dimensions of 6.8$\times$6.8 arcmin on the sky (they overlapped by about 2 arcmin in the E-W direction), with a plate scale of 0.2''/pix. 
For each field, four dithered images were acquired with each filter, each with exposure times of 94\,s (on-band), 90\,s (off-band), and 10\,s ({\it r}). The images were reduced and calibrated with the normal procedures of the ESO pipeline. That is, reduced images are bias-subtracted, flat-fielded, and astrometrically calibrated. The astrometric precision is  0.2 arcsec.

Each set of four images was combined using IRAF{\footnote{IRAF is distributed by the National Optical Astronomy Observatories, which is operated by the Association of Universities for Research in Astronomy, Inc., under contract to the National Science Foundation.}  routines, allowing for cosmic ray removal. The resulting images are  equivalent to a 6.2 min exposure for the [O {\sc iii}] on-band filter, 6 min for the off-band filter, and 40 s for the {\it r} filter (these images will be available at the CDS). The on- and off-band combined images were subtracted (after background subtraction and normalization) to easily detect [O {\sc iii}]~5007 emission objects.

No standard stars were observed for the photometric calibration of the images. Nevertheless in \S 3 we show that a rough calibration of the stellar magnitudes can be performed using   some known stars from the field, and the [O {\sc iii}] on-band images can be calibrated using  our spectroscopy.
\bigskip

The continuum-subtracted [O {\sc iii}] 5007 images show numerous [O {\sc iii}] 5007-emitting objects.   The 7 PN candidates and the 10 HII regions reported by Richer \& McCall (1992) were recovered.  At least four of the large shells studied by Hunter et al. (1993) are clearly visible; the others are not, probably because they are low excitation nebulae that do not emit in [O {\sc iii}].  Several of  the HII regions reported by Hodge (1969) were also identified. In addition, a number of small diffuse HII regions exist,  as well as many unresolved  emitting objects. Some of them coincide with the H$\alpha$ emitting objects in the list from Bresolin et al. (1993), but the coordinates provided by these authors have uncertainties larger than 1$''$, and sometimes it is difficult to identify  their objects with ours.   Our objects are presented in Tables 1 and 2, where  we have included a column named ``other ID''  listing the names given by other authors for the same object.
\bigskip

One of the most important issues in studying  extragalactic PNe is to distinguish true PNe  from compact HII regions and, of course, to avoid contamination with background objects. To identify PNe in our  imaging, we  used the criteria described below. Criteria (a) and (b) are very similar to those employed by Soffner et al. (1996) in their work on NGC 300.
\smallskip

\noindent{\bf a. Object size} 

At the distance of NGC~3109 (about 1.33 Mpc), one arcsec corresponds to  6.4 pc. Given our pixel size of 0.2 ''/pix and our FWHM of about 3.2 pix for stellar objects, PNe (which typically have diameters smaller than 1 pc) should appear as point-like objects. Thus, we selected the unresolved [O {\sc iii}]-emitting objects. These nebulae  can be either PN candidates or unresolved  HII regions. 
\smallskip

\noindent{\bf b. The central star} 

HII regions and PNe with significant [O {\sc iii}] 5007 emission are ionized by stars with T$_{\rm eff} \geq$ 35,000 K (e.g.,  Stasi\'nska 1990). In the case of HII regions, this implies a central star of spectral type O8V or earlier (or O7 if the luminosity class is III or Ia). Typically, such stars have absolute visual magnitudes of M$_{\rm V} \sim -4.3$ to $-4.6$ or brighter (Martins et al. 2005; Vacca et al. 1996). On the other hand, the central stars of galactic PNe with T$_{\rm eff} \geq$ 35,000 K typically have M$_{\rm V} \sim -2.0$ (M\'endez et al. 1992), which is more than 2 mag  fainter than the ionizing stars of HII regions. Adopting a distance modulus of 25.6\,mag for NGC 3109  implies an apparent visual magnitude, V, brighter than about 21.3 mag for the ionizing stars in HII regions and fainter than about 23.3 mag for the central stars of PNe.

By measuring the visual magnitudes of the central stars of unresolved emitting objects, we can distinguish compact HII regions from PN candidates (see next section). Of course, it is possible that a foreground or background star may appear projected on a PN candidate. In that case, our criterion would classify the object as an HII region rather than a PN, and spectroscopy is needed to classify the object properly. 
\smallskip

\noindent{\bf c. The {\it r} images}

The {\it r} filter has its maximum transmission at 6530 \AA\ and an FWHM of 810 \AA. It therefore includes the stellar flux and the nebular lines in this wavelength range.
In particular, it includes H$\alpha$. Thus, we expect the [O {\sc iii}]-emitting objects to appear in the {\it r} images as well. Most of the objects do appear, in which cases  they are most probably not background emission-line objects. However, requiring detection in both [O {\sc iii}] and {\it r} discriminates against  very faint (e.g., Magrini et al. 2001) but highly excited PNe that should have no stellar emission and for which the 5007/H$\alpha$ ratio is large. Therefore, we will also consider with caution those candidates with faint [O {\sc iii}], but which do not appear in the {\it r} image.
\smallskip

\noindent {\bf d. The luminosity of recombination lines}

In addition to the previous criteria,  it should be taken into account that, often in HII regions, there is more than one ionizing star and these stars 
produce a larger number of ionizing photons, Q$_0$, due to their larger radius.  Typically log Q$_0 \geq$ 48.0 (s$^{-1}$) for HII regions (Vacca et al. 1996) versus log Q$_0 \sim$ 47.5 for PNe (using evolutionary tracks by Bl\"ocker 1995). Optically thick compact HII regions should then be brighter than optically thick PNe in the hydrogen recombination lines. Furthermore,  many PNe in the Milky Way are known to be optically thin, further weakening these lines. The H$\alpha$ or H$\beta$ fluxes (for instance) can be used for this  criterion.
\smallskip

\noindent{\bf e. The ionization degree or excitation class}

It is expected that, statistically, PNe will be more highly excited than HII regions as a result of containing hotter ionizing stars. If a nebula shows He {\sc ii} 4686 emission whose intensity exceeds a few percent of H$\beta$, it is unquestionably  a PN. Also, intensity ratios  [O {\sc iii}] 5007 / H$\beta$ larger than about 4\ are found only in PNe. However, this criterion discriminates against PNe whose central stars have low T$_{\rm eff}$ (usually young  PNe). Low-excitation PNe cannot be distinguished from HII regions on the basis of only their spectroscopic line ratios.   
\bigskip

 \noindent From the above  criteria,  {\it a} and {\it b} are clearly the least ambiguous in distinguishing between compact HII regions and PN candidates, when only imaging is available. The outcomes of our imaging analysis, upheld with the [O {\sc iii}] fluxes from our spectroscopy,  are presented in the following sections.

\section{Nebular and stellar magnitudes}

For all of our [O {\sc iii}] emitting objects, we have determined several instrumental magnitudes: the nebular magnitude m$_{\rm i}$(5007) from the continuum-subtracted on-band image, the stellar magnitude m$_{\rm i}$(star) from the off-band image, and m$_{\rm i}$({\it r}) from the {\it r} image.  Magnitudes were determined using the `apphot' package in IRAF. Aperture photometry was performed by integrating the flux within a 5 pixel radius (equivalent to 1 arcsec) centered on the stellar position.  The background was determined from an annulus 10-13 pixels from the star and subtracted from the  flux. Special care was taken in crowded fields. Instrumental magnitudes for each object are listed in Tables 1 and 2 for the PN candidates and the HII regions, respectively.
 
\subsection{Nebular [O {\sc iii}] 5007 magnitudes}
The nebular magnitude, m$_{\rm i}$(5007), was measured from the continuum-subtracted [O {\sc iii}] image, thereby excluding the emission from the central star.  The aperture employed  includes all the flux for the nebulae with radii smaller than 1 arcsec, but it contains only a fraction of the flux for very extended objects.   We find that the measured magnitudes for unresolved objects are well-correlated with the  [O {\sc iii}]\,5007 fluxes measured from our spectroscopy through the relation:

log F(5007) = $-$0.40 m$_{\rm i}$(5007) $-$ 8.32 \ .  \hskip 2cm  (1)

\noindent This relation (shown in Fig. 1) has  a small dispersion ($\leq$0.05 mag) and a large correlation coefficient R$^2$=0.997.  This implies that the  FORS1 pre-imaging,  reduced through the ESO pipeline (\S 2), is very  consistent with the calibrated spectrophotometry. Thus, we are confident that the differences in instrumental magnitudes do represent flux ratios.

\subsection{Stellar magnitudes}

Apparent visual magnitudes for the central stars, m$_{\rm i}$(star), were measured in the off-band frame. A rough photometric calibration using the same procedure as employed by Richer \& McCall (1992) was used to relate these instrumental magnitudes to V magnitudes.  That is, some stars from the list of Demers et al.  (1985; hereafter DKI) with measured B and V magnitudes  were used as local standards. Certainly, a calibration based upon observations of photometric standards would have been more appropriate for a photometric calibration; but, as said in \S 2, our images were acquired as `pre-imaging'  for spectroscopy, so no standards were observed. 

 Our off-band instrumental magnitudes follow a linear relation  with  the V magnitudes from DKI, V(DKI), given by the expression:

V(DKI)=0.99 m$_{\rm i}$(star) + 4.38 \ .    \hskip 3cm  (2)

\noindent This relation is plotted in Fig.1. It has a dispersion $\leq$0.15 mag and a correlation coefficient R$^2$=0.989.
The good correlation of the two sets of magnitudes again indicates that differences in  m$_{\rm i}$(star) indeed represent flux ratios, which allows us to assign a visual magnitude, V(DKI), to the central stars of our HII regions and candidate PNe. 

We also measured the instrumental {\it r} magnitude, m$_{\rm i}$({\it r}). As already noted, m$_{\rm i}$({\it r}) includes the stellar flux and the nebular lines in the {\it r} filter, particularly H$\alpha$.  Consequently, objects with bright m$_{\rm i}$({\it r}) magnitudes have very intense H$\alpha$ or a bright star. 

\subsection{PNe and HII regions in NGC 3109}

We have classified the diffuse and extended emission nebulae as HII regions. In the case of the compact objects, we separate HII regions from PN candidates using the magnitudes of their ionizing stars.  Considering that the visual magnitudes of the ionizing stars in HII regions should be brighter than about 21.4 mag (\S 2, b), equivalent to 17.2 mag in our system, we  consider all objects with m$_{\rm i}$(star) brighter than 18 mag as HII regions.  This limiting magnitude allows for an uncertainty of 0.3  mag in individual instrumental magnitudes (twice the dispersion of Eq. (2)) and an extinction of up to 0.5 mag, a value that amply includes the small color excess of E(B-V)= 0.04 mag (equivalent to A$_{\rm V}$ = 0.13 mag, see \S 4) reported by Burstein \& Heiles (1984) in the direction of NGC 3109 and any plausible internal extinction in the HII regions.  We note that some ionizing stars, in both HII regions and PNe, could be binaries, in which case m$_{\rm i}$(star) and V(DKI) represent the fluxes of both stars.  Considering the possibility of binarity, our limiting magnitude for separating PNe from HII regions is then quite conservative.
 
In Fig. 2, we present the color index m$_{\rm i}$({\it r})$-$m$_{\rm i}$(star) plotted as a function of m$_{\rm i}$(star) for HII regions (objects with m$_{\rm i}$(star) $\leq$ 18) and PN candidates (with m$_{\rm i}$(star) $>$ 18).  The colors of  HII regions  span the interval from $-$1.5  to 0.8 mag. For PN candidates  the colors  are bluer (more negative) than $-$1.3.  The two types of nebulae are separated very nicely in this figure. This separation occurs because the color index m$_{\rm i}$({\it r})$-$m$_{\rm i}$(star) corresponds roughly to the flux ratios (F(H$\alpha$) + F$_{r}$(star))/F$_{i}$(star), which will be large for PNe while, at the same time, their central stars are faint.

\begin{figure}
\includegraphics[width=9cm]{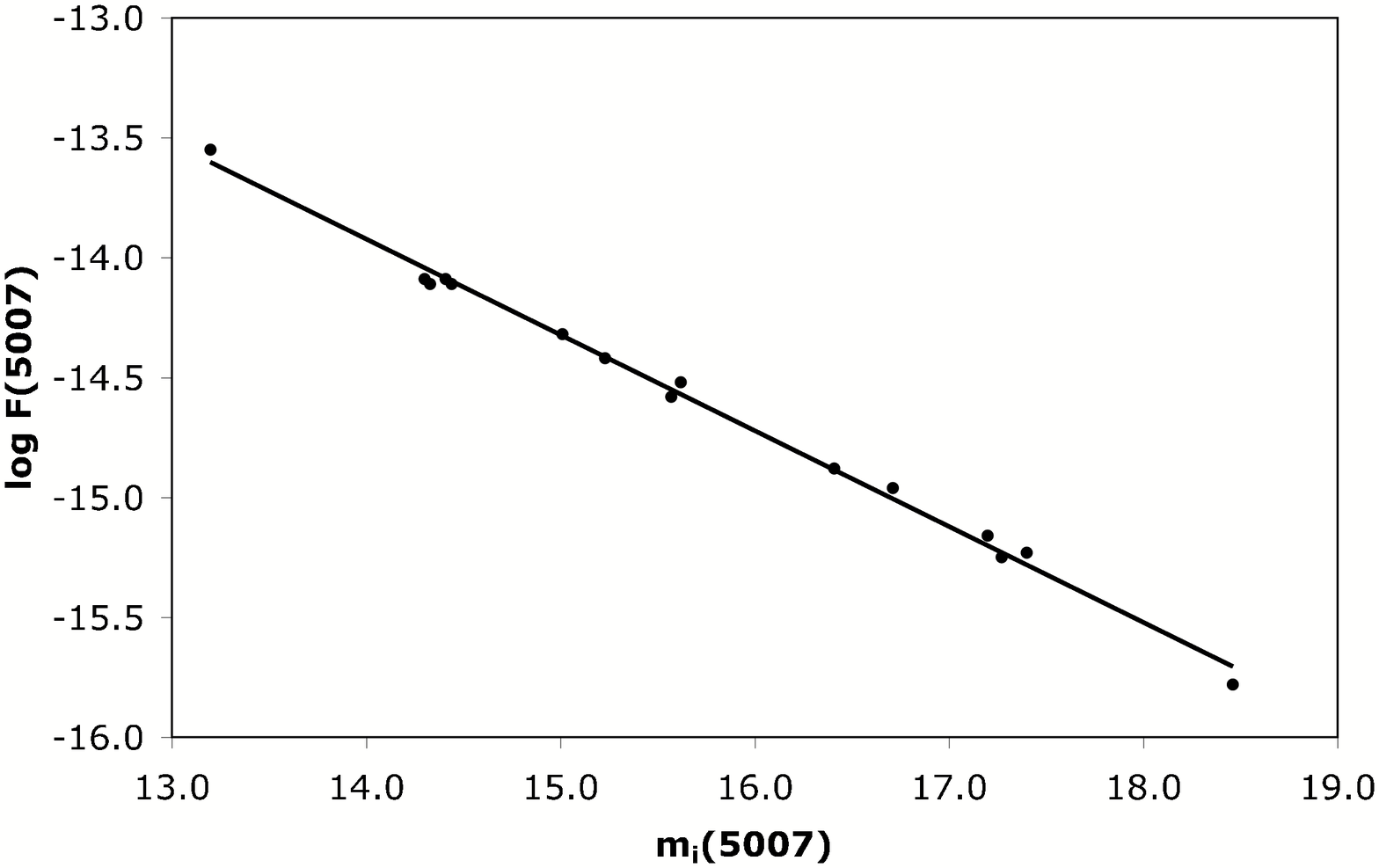} 
 \includegraphics[width=9cm]{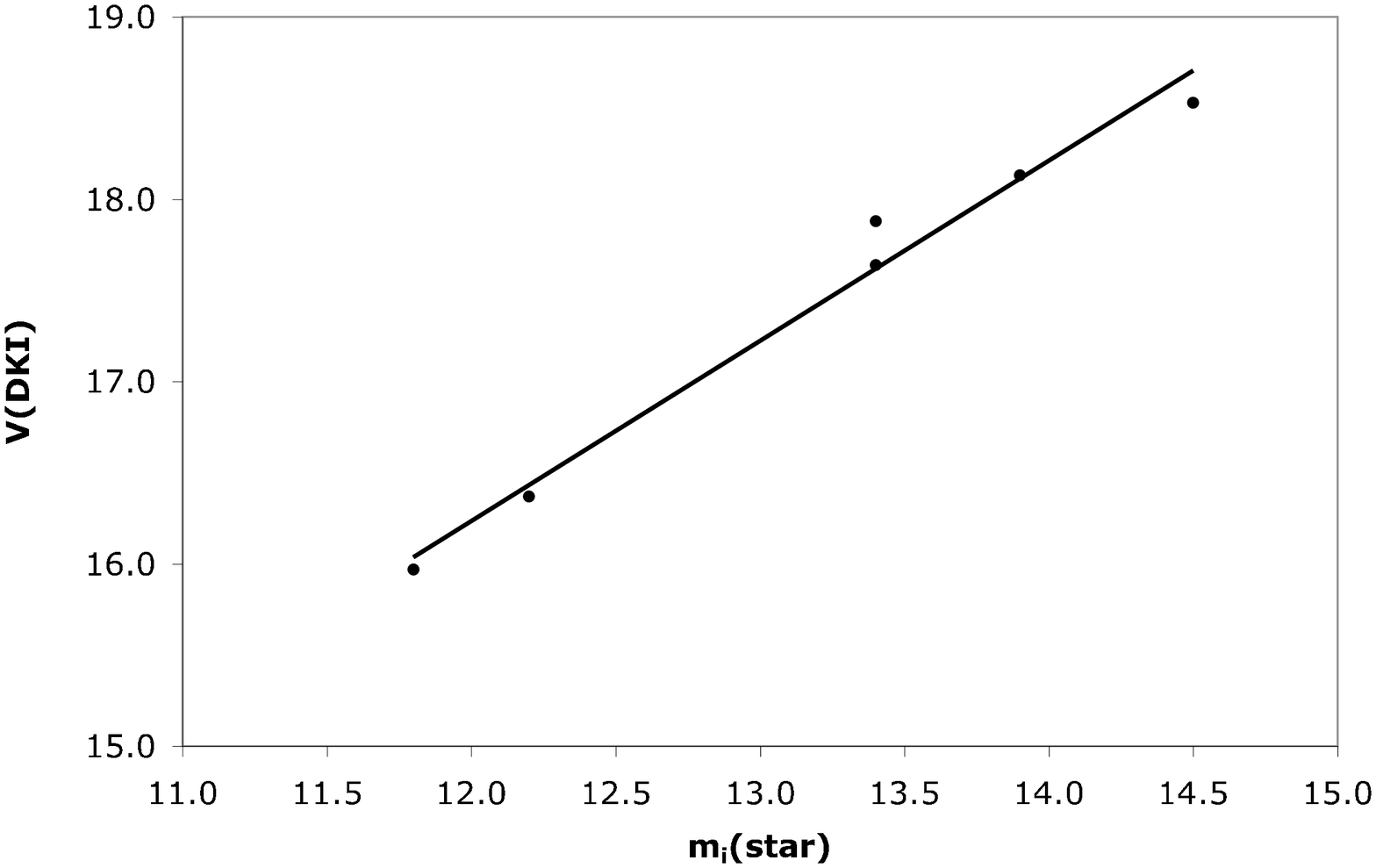}
\caption{The relations between our instrumental m$_{\rm i}$(5007) and the spectroscopic [O {\sc iii}] 5007 flux, F(5007), (upper panel) and between m$_{\rm i}$(star) and V(DKI) for some stars (lower panel).\label{fig1}}
\end{figure}

\begin{figure}
\includegraphics[width=9cm]{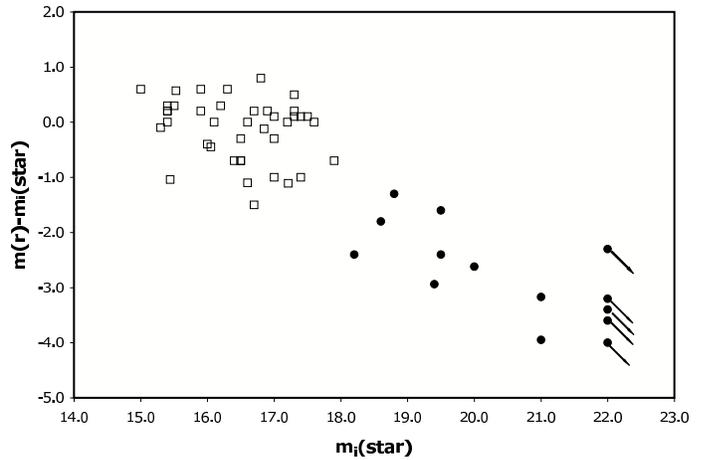}
\caption{The color index  m$_{\rm i}$({\it r})$-$m$_{\rm i}$(star)  plotted against m$_{\rm i}$(star).  Open squares represent HII regions (objects with m$_{\rm i}({\mathrm star}) \leq$ 18 mag). Black dots represent PN candidates.  Since {\it r} includes the flux from the star and H$\alpha$, the color index is proportional  to the flux ratio (F(H$\alpha$)+F$_r$(star))/F$_i$(star).  Thus, objects with a very faint central star (e.g., PNe) appear in the lower right corner.  The values m$_{\rm i}$(star)= 22 are upper limits. For these objects, the star was not detected. The arrows show the displacement of dots  for stars with fainter m$_{\it r}$. \label{fig2}}
\end{figure}

\bigskip

In total, we detected 20 PN candidates. Their characteristics are listed in Table 1, where we include  their coordinates (columns 2 and 3) and magnitudes  m$_{\rm i}$(5007), m$_{\rm i}$(star), V(DKI), and m$_{\rm i}$({\it r}) (columns 4 to 7, respectively). Column 8 presents log F(5007), in erg cm$^{-2}$ s$^{-1}$, from our subsequent spectroscopy (Pe\~na et al., in preparation), while column 9 lists the value of log F(5007) calculated from Eq. (1). In column 10, any previous identifications are listed. As already reported in Pe\~na et al. (2006), we confirm as PNe only two of the seven PN candidates found by Richer \& McCall (1992).  

\begin{table*}
\caption{ Positions and characteristics of  PN candidates in NGC~3109$^{a}$}
\label{table:1}
\centering
\begin{tabular}{lcccrrrccl}
\hline \hline \\
Name&	R.A.$^{b}$ & Dec$^{b}$ &	m$_{\rm i}$(5007) & m$_{\rm i}$(star)$^{c}$ &  V(DKI)  & m$_{\rm i}$({\it r})$^{c}$& log$^{d}$ &  log$^{e}$ &other ID$^{f}$\\
           &          &        &        &       &       &  & F(5007) &F(5007)   \\
\hline 						
PN 1&	10:02:41.57& -26:08:40.7&	19.0&	$>$ 22 &	$>$ 25& $>$ 22.5 & & $-$15.91 & \\
PN 2&      10:02:44.70 & -26:11:32.4 &	19.7:&	21.0: & 25.1:& 17.8 & & $-$16.21  & \\
PN 3&	10:02:49.15& -26:09:10.1&	16.4&	$>$ 22&	$>$ 25& 18.8 & $-$14.88& $-$14.89 & \\
PN 4&	10:02:50.06& -26:08:10.0&	17.3&	$>$ 22&	$>$ 25& $>$ 22.5 & $-$15.25&$-$15.23&  \\
PN 5&       10:02:51.60& -26:10:19.7&	18.7&	21.0: & 25.1: & 17.1&  & $-$15.81 \\
PN 6&	10:02:57.20& -26:08:55.8&	18.7&	19.5 & 23.6 & 17.9 &  & $-$15.81&\\
PN 7&	10:02:58.20& -26:08:45.8&	14.4&	18.2&	22.4& 15.8 &$-$14.09&$-$14.09& PN1 RMc \\
PN 8&	10:02:59.52& -26:08:13.9& 	18.9&	20.0& 24.1 & 17.4 &  & $-$15.89 &\\
PN 9&	10:03:02.66& -26:08:50.3&	16.4&	$>$ 22 &	$>$ 25& 19.7 &  & $-$14.88    & \\
PN 10&	10:03:03.91& -26:09:42.3&	15.4&	$>$ 22 &	$>$ 25& 18.4  &  & $-$14.48 & PN6 RMc\\
PN 11&	10:03:05.04& -26:10:54.2&	15.6&	$>$ 22 & $>$ 25& 18.0 & $-$14.52& $-$14.57 & \\
PN 12&	10:03:07.99 &  -26:07:57.8&	18.8&	19.40& 23.5 & 16.5&  & $-$15.21   \\
PN 13&	10:03:08.52& -26:09:07.7&	17.2&	$>$ 22&	$>$ 25& $>$ 22.5 & &  $-$15.15& \\
PN 14&	10:03:10.34& -26:08:06.1&	18.5&	19.5&	23.2& 17.1 & $-$15.78 & $-$15.71 & \\
PN 15&	10:03:13.32& -26:10:20.7&	19.3&	$>$ 22&	$>$ 25& $>$ 22.5 &  & $-$16.04 & \\
PN 16&	10:03:18.67& -26:09:58.8&	17.0&	$>$ 22&	$>$ 25& 18.6 &  & $-$15.13 &	 \\
PN 17&	10:03:18.74& -26:10:05.4&	16.7&	18.8&	23.0& 17.5 & $-$14.96 & $-$15.01  \\
PN 18&	10:03:20.74& -26:09:00.2&	17.5&	$>$ 22 &	$>$ 25& $>$ 22.5 &  & $-$15.32 & \\
PN 19&	10:03:21.22& -26:11:20.9&	19.1&	$>$ 22 &	$>$ 25& $>$ 22.5 &  & $-$15.98 & \\
PN 20&	10:03:25.51& -26:09:07.2&	17.0&	18.6&	22.8& 16.8 &  & $-$15.13 &  \\
\hline 
\multicolumn{6}{l}{(a) All these objects are  unresolved.}\\
\multicolumn{6}{l}{(b) Coordinates for equinox 2000.}\\
\multicolumn{10}{l}{(c)  Objects with m$_{\rm i} >$ 22 are not detected in the off-band image. Objects with m$_{\rm i} ({\it r}) > $ 22.5 are not detected in the {\it r} image.}\\
\multicolumn{6}{l}{(d) Flux from spectroscopy, in erg cm$^{-2}$ s$^{-1}$.}\\
\multicolumn{6}{l}{(e) Flux calculated from Eq. (1), in erg cm$^{-2}$ s$^{-1}$.}\\
\multicolumn{6}{l}{(f) RMc: Richer \& McCall 1992.}
\end{tabular}
\end{table*}

Table 2 lists the characteristics of the HII regions. We present coordinates, diameters, magnitudes, and fluxes  for  40  nebulae found in our continuum-subtracted [O {\sc iii}] image. This is not an exhaustive list of HII regions in NGC~3109, but it includes a large fraction of the [O {\sc iii}]~5007 emitting regions. The very low excitation nebulae without 5007 emission are obviously not in this list, and some detected but very faint HII regions and filaments were not included.  The m$_{\rm i}$(star) and  V(DKI) magnitudes measured for the central star of the HII regions (when there is an obvious central star) are presented in columns 6 and 7, respectively. In column 8 we list m$_{\rm i}$({\it r}). The  [O {\sc iii}] 5007 fluxes, F(5007), from spectroscopy and calculated via Eq. (1)  are listed  in columns 9 and 10, respectively. The other columns contain a description of the morphology (diffuse, compact, extended, complex, unresolved), a rough estimate of the stellar spectral type (see below), and previous identifications of the HII regions (columns 11-13, respectively).

In Fig. 3 we present our [O {\sc iii}] 5007  image of NGC~3109.  In this figure, we show the distribution of  PNe from Table 1 and  compact HII regions from Table 2.   Our objects are spread throughout the galaxy, and it is evident that  HII regions are rather concentrated towards the  edge-on disk of the galaxy, while PNe are seen also above and below this structure, consistent with their belonging to an older stellar population.

\begin{figure*}
\includegraphics[width=\textwidth]{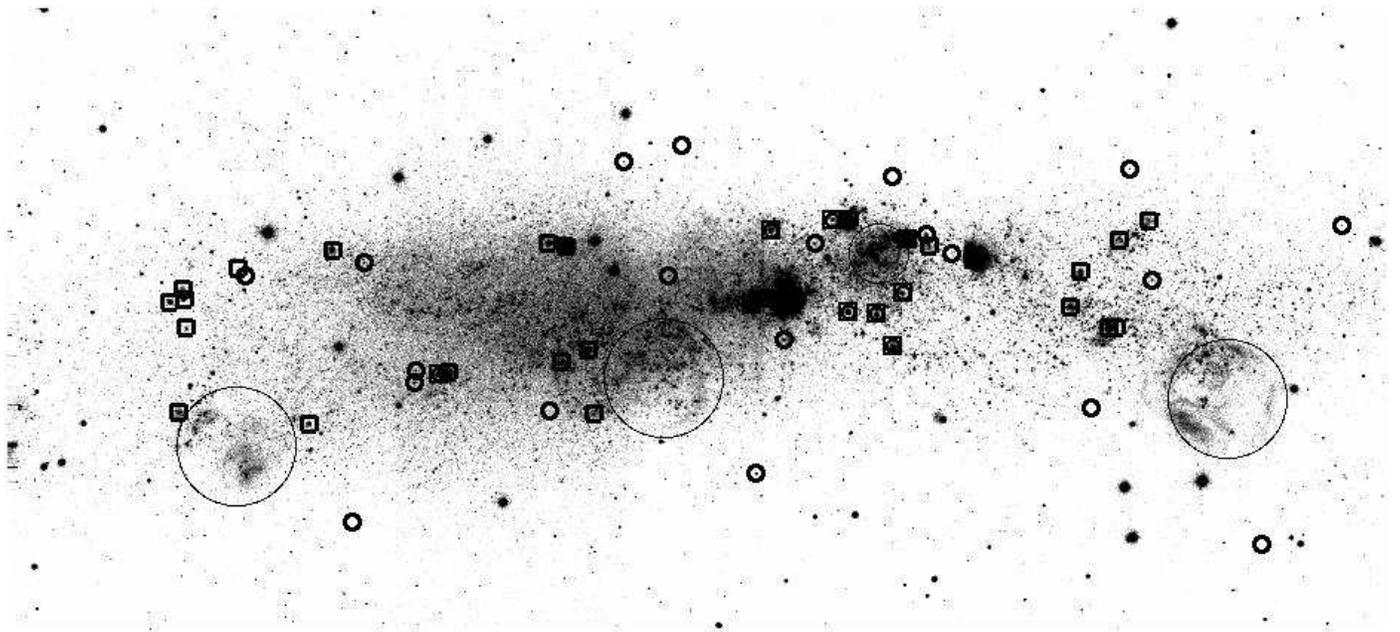}
\caption{Our  [O {\sc iii}] on-band image, showing the whole galaxy. North is up, east is left. The full size of the image is  16.7 $\times$ 5.3  arcmin. PN candidates from Table 1 are marked with a circle. Compact HII regions from Table 2 are marked with a square (squares with an inner circle indicate the objects reported by Richer \& McCall 1992). The HII regions are rather concentrated towards the edge-on disk of the galaxy, while PNe spread over a wider area. The large circles, from left to right, show the shells \#1, 3, 5, and 9 studied by Hunter et al. (1993). \label{fig3}}
\end{figure*}

\subsection{Individual PNe}

\noindent{\bf a. The brightest PN}

According to our selection criteria, our PN\,7 and PN\,10 are the only  bona fide PN candidates from the list reported by Richer \& McCall (corresponding to their PN\,1 and PN\,6, respectively). PN\,10 is a bright PN with no stellar continuum, but PN\,7 has the brightest central star among our PN candidates, with V(DKI)=22.4 mag, i.e., brighter than expected for a single star (\S 2, criterion b). This nebula is also the brightest  in [O {\sc iii}]\,5007, one magnitude brighter than the next brightest object.  We included it in the PN group  due to its spectroscopy, which shows a very high 5007/H$\beta$ ratio (Pe\~na et al., in preparation). 
In addition, it shows a [O {\sc iii}] 5007 line width, HWHM = 13 km s$^{-1}$ (Richer 2006), which is significantly greater than the values of HII regions. For comparison, the average
[O {\sc iii}] 5007 line width (HWHM) of the five PN candidates from Richer \&
McCall (1992) that we find are compact H II regions is 11 km s$^{-1}$ (Richer 2006; Richer et al. in preparation).
 Also in \S 4 we present further arguments for including it among the PN sample.

\noindent {\bf b. The PNe with no or very faint {\it r} emission}

Six PN candidates in our sample were not detected in our {\it r} image. They are PN\,1, PN\,4, PN\,13, PN\,15, PN\,18, and PN\,19. 
We determined an instrumental upper limit of 22.5 mag for their m$_{\rm i}$({\it r}). Several of them are also very faint in [O {\sc iii}]. 
As mentioned in \S2 (criterion c), these objects are considered PN candidates but some of them  could be background sources.
Most probably, though, they are high excitation PNe where [O {\sc iii}] 5007 is much stronger than H$\alpha$.
From our spectroscopy, we have verified that at least PN\,4 is a true PN (Pe\~na et al., in preparation).

\subsection{The stellar spectral types in HII regions}

On the V(DKI) magnitude scale, the central stars of HII regions show values in the range from 22.1 to 19.5, and a spectral type can be assigned to each them by adopting an absolute magnitude scale for massive stars (for instance Vacca et al. 1996 or Martins et al. 2005) and a distance modulus. We  chose the distance modulus, m-M= 25.62 mag (Minniti et al. 1999).  This exercise was  done only for those relatively small HII regions where there is an obvious  central star. Extended nebulae (complex regions and loops) appear related to  stellar clumps and it is very difficult to decide which ones are the ionizing stars; therefore for them we have not tried to identify the ionizing stars.  Another problem is the assumption that these stars are single, which likely biases our classifications to earlier spectral types.  In any case, these spectral types (column 9, Table 2) are  very rough approximations that should be considered merely indicative.  In addition, no reddening has been  assumed nor has any account been taken of any possible magnitude offset of the DKI calibration (Bresolin et al. 1993 claimed a difference of about 0.3 - 0.4 mag).  From this rough approximation we detect a few giants and supergiant stars (spectral type O III or O Ia), five  main sequence stars between O4 and O6, five O6-O7, and 16 stars later than O7.
 
\addtocounter{table}{1}

\section{The luminosity function for PNe}

From our [O {\sc iii}] fluxes, F(5007) (Eq. 1; column 9, Table 1), an apparent magnitude m$_{5007}$ can be obtained and the [O {\sc iii}] 5007 luminosity function (PNLF)  can be constructed.
In Fig. 4, we present the cumulative luminosity function for our PN 
candidates (including PN\,7, see below).  We chose this representation 
because of the low number of PN candidates and because the peak 
magnitude of the luminosity function is poorly constrained at the
metallicity of NGC 3109 (Ciardullo et al. 2002).  Therefore, we must 
fit the cumulative luminosity function using apparent magnitudes, 
instead of the usual absolute magnitudes.  We convert  the usual 
luminosity function

N(M) $\propto \exp(0.307 M_{5007}) ~ (1-\exp(3(M^*_{5007}-M_{5007})))$
 
\noindent to a luminosity function in apparent magnitudes

N(m) = N $\exp(-0.307 \mu) ~\exp(0.307 m_{5007})~ (1-\exp(3(m^*_{5007} - m_{5007}))$\ ,

\noindent where N is a normalization constant, $\mu$ is the apparent 
distance modulus, $\mu = 5\log \mathrm d -5 + A_{5007}$, $M^*_{5007}$ 
and $m^*_{5007}$ are the absolute and apparent peak magnitudes of the 
luminosity function, respectively, the latter defined via

m$_{5007}$ = $-2.5 \log$ F(5007) -13.74

\noindent (Allen 1973; Jacoby 1989), and $A_{5007}$ is the extinction 
at 5007 \AA\ towards NGC 3109.  For the extinction, we adopted the 
aforementioned color excess, $E(B-V) = 0.04$\,mag, and the Fitzpatrick 
(1999) reddening law, parametrized with a total-to-selective 
extinction of 3.041.  This parametrization delivers a true ratio of 
total-to-selective extinction of 3.07 when integrated over the 
spectrum of Vega (McCall 2004), which is the average value for 
the diffuse component of the interstellar medium of the Milky Way 
(McCall \& Armour 2000).  The extinction towards NGC 3109 then 
amounts to $A_{5007} = 0.13$ mag.  Our fit to the cumulative luminosity function 
returns the product $N\exp(-0.307 \mu)$ and $m^*_{5007}$ as fitting parameters.  
Although this allows us to derive a normalization constant without knowledge 
of the peak magnitude of the luminosity function, it obviously 
precludes using our peak magnitude to derive a distance to NGC 3109, as is usually done.

Since the brightest PN (PN\,7 in Table 1, see \S 3.4) is 1\,mag brighter than the next brightest 
object, we fit the cumulative luminosity function while both 
including and excluding this object.  Both fits to the luminosity 
function are similarly good, so the luminosity function provides no 
grounds for excluding this object as a PN.  The parameters for these 
fits are given in Table 3, where we also provide similar fits to 
the cumulative luminosity functions for PNe in the Magellanic Clouds 
and M31 for comparison.  Our apparent peak magnitudes for M31 and the 
Magellanic Clouds are within 0.12\,mag of those found in the original 
studies (see Ciardullo et al. 1989 for M31), a difference that falls 
within the uncertainties in all of these studies, including ours.

\begin{table}
\caption{ Fit for the cumulative PNLF}
\begin{tabular}{lccc}
\hline \hline 
fit           & $m^*_{5007}$ & $N\exp(-0.307 \mu)$      \\
\hline
including PN\,7 &  21.18       & $2.62\times 10^{-3}$    \\
excluding PN\,7 &  21.82       & $2.77\times 10^{-3}$    \\
LMC$^{\mathrm a}$& 14.23     & 0.287                   \\
SMC$^{\mathrm a}$& 14.82     & $9.31\times 10^{-2}$    \\
M31$^{\mathrm b}$& 20.20     & 0.150  \\
\hline\hline \\
\multicolumn{3}{l}{(a) Based upon the sample of Jacoby et al. (1990) and} \\
 \multicolumn{3}{l}{considering the 
same limiting magnitude, 15.9\,mag, for fitting.}\\
\multicolumn{3}{l}{(b) Based upon the sample of Ciardullo et al. (2002) and}\\ 
\multicolumn{3}{l}{fitting to the 
first three magnitudes of the luminosity function.}\\
\end{tabular}
\end{table}

Considering the brightest PN candidate as a bona fide PN, we can use 
the true distance modulus of 25.62\,mag (Minniti et al. 1999) and the extinction A$_{5007}$=0.13 to 
derive a peak absolute magnitude of the luminosity function of $M^*_{5007} = -4.57$
\,mag for NGC 3109. This value is surprisingly  bright at the metallicity of this galaxy, compared 
with the value of $-4.48$ mag suggested by Ciardullo et al. (2002) from studies of several galaxies of higher metallicity.  On the other hand, excluding the brightest PN, our fit to the 
luminosity function implies a peak absolute magnitude of only $-3.93$\,mag, 
a value that is  faint compared to the values of about $-4.08$ mag expected by Ciardullo et al. (2002) for galaxies with metallicities 12 +log O/H $\sim$ 8.0.  
Obviously, one of the benefits of observing low metallicity galaxies 
like NGC 3109 will be a better understanding of the absolute 
luminosities of their planetary nebulae.

Based upon the normalization factors in Table 3, NGC 3109 has about 
the expected number of PNe.  Adopting a true distance modulus  $\mu_0$=18.50\,mag
and A$_{5007}$ = 0.36\,mag for the LMC (Jacoby et al. 1990)  
and $\mu_0$=25.62\,mag and A$_{5007}$= 0.13\,mag for NGC 3109,  we find $N = 
93.85$ and $N = 7.11$ for the LMC and NGC 3109, respectively.  If we 
adopt the absolute $B$ magnitudes from Richer \& McCall (1995), but 
correct them to the distance moduli already mentioned, the luminosity ratio is 11.69, 
very similar to the ratio of $N$ values, 13.20.  Therefore, NGC 3109 
has about the same number of PNe per unit luminosity as the LMC. 
This is expected, since the colors of the LMC and NGC 3109 are similar 
(Buzzoni et al. 2006). 

 In Fig. 4, we overlay the LMC's luminosity 
function, scaled by $1/13.2$.   Both fits (NGC 3109's and LMC's) show excellent agreement over the first 1.5$-$2.0\,mag, the extent over which the LMC's luminosity function is 
believed to be complete (e.g., Richer et al. 1997).  Since the normalization of the two fits to the luminosity function are similar (Table 2), regardless of whether PN\,7 is considered, the conclusion that NGC 3109 has about the number of PNe expected is independent of whether PN\,7 may or may not  be a true PN. 

\bigskip

\begin{figure}
\includegraphics[angle=-90,width=9cm]{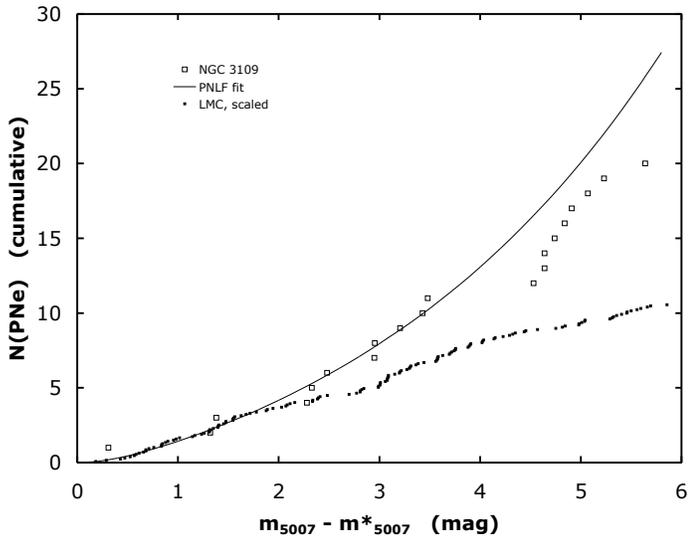}
\caption{Cumulative planetary nebula luminosity function from the data in Table 1. The fit to the cumulative PNLF returns the product $N\exp(-0.307 \mu)$ and $m^*_{5007}$ (see text and Table 3). The LMC's luminosity function, scaled by a factor of 1/13.2 is also presented.\label{fig4}}
\end{figure}

\section{Conclusions}

From [O {\sc iii}]~5007 on-band, off-band, and {\it r} imaging, we detected a large number of extended and compact emitting nebulae in NGC~3109. 
We have defined criteria to distinguish between PN candidates and compact HII regions, which allow us to identify 20 PN candidates, for which we present coordinates, [O {\sc iii}] 5007 magnitudes, and stellar magnitudes, when the central star is detected.   We also present a catalog of  40 HII regions (extended and compact) that includes coordinates, diameters, [O {\sc iii}] 5007 magnitudes, and instrumental magnitudes for the central stars. Apparent V magnitudes  have been estimated from a rough photometric calibration of the instrumental magnitudes. Based upon these V magnitudes, we assign a rough spectral class to the ionizing stars of HII regions.
\medskip

The distribution of  HII regions  is concentrated towards the edge-on disk of the galaxy, while PN distribution is more extended, covering a wider area, above and below the disk. This is consistent with the PNe belonging to an older stellar population.
\smallskip

The cumulative PNLF for [O {\sc iii}] 5007 magnitudes was constructed and a fit to this function was calculated. Comparing this fit with a fit to the available data for the LMC shows that NGC 3109 has about the same number of PNe per unit luminosity as the LMC. By assuming a true distance modulus of 25.62 mag and an extinction A$_{5007}$=0.13 mag for NGC 3109, a peak value of $-$4.57\,mag is derived for the PNLF.  This value is high compared both with the canonical $-$4.48\,mag found for galaxies with higher metallicity and with the peak magnitude expected at the metallicity of NGC 3109.  By studying more galaxies with metallicities similar to NGC 3109's, a better understanding of the absolute luminosities of low metallicity PNe will be acquired.

\begin{acknowledgements}
M. Pe\~na is grateful to the Departamento de Astronom{\'\i}a, Universidad de Chile for hospitality during a sabbatical stay
when part of this work was performed. M.P. gratefully acknowledges financial support from FONDAP-Chile and DGAPA-UNAM. We thank the referee, Dr. Mario Perinotto, for his constructive comments. This work received financial support from grants \#43121 (CONACyT-Mexico), IN-108506-2, IN-108406-2, IN-112103, and IN-114805 (DGAPA-UNAM).
\end{acknowledgements}

\clearpage
\longtabL{2}{
\begin{landscape}
\begin{longtable}{cccrcccccrrcc}
\caption{Positions and characteristics of  HII regions in NGC~3109 \label{tbl-2}}\\
\hline\hline
Name & R.A. (2000) & Dec (2000) & $\phi$('') & m$_{\rm i}$(5007) &
m$_{\rm i}$(star) & V(DKI)$^{a}$ &m(r) &  log F$^{b}_{5007}$ & log F$^{b}_{5007}$ & description$^{c}$ & S. T. & other ID$^{d}$\\
\hline 
\endfirsthead
\caption{continued.}\\
\hline\hline
Name & R.A. (2000) & Dec (2000) & $\phi$('') & m$_{\rm i}$(5007) &
m$_{\rm i}$(star) & V$_{DKI}^{a}$ &m(r) &  log F$^{b}_{5007}$ & log F$^{b}_{5007}$ & description$^{c}$ & S. T. & other ID$^{d}$\\
\hline
\endhead
\hline
HII 1   & 10:02:46.50 & -26:10:10.0 & $>$ 60& ---  & ---   & ---    & --- &      & --- &cmplx loop&  clump& Ho19; Hu9; F3H1,F3H2,.. \\
HII 2   & 10:02:49.35 & -26:08:38.1 &      5.7 &  18.6: &16.9   &21.0 & 16.7 &    & $-$15.8:   & diff  & O7-O8 V& F2H4\\
HII 3   & 10:02:50.50 & -26:08:48.5 &      5.2 &  18.8: &16.2   &20.4    & 16.5 &      &  $-$15.9: & diff & O4-O5 V & F2H5 \\
HII 4   & 10:02:50.66 & -26:09:35.9 &      1.1 &14.7 &  16.1 &  20.2& 15.6 &     &$-$14.21& comp & O4-O5 V &   near F2H7   \\       
HII 5   & 10:02:50.89 & -26:09:35.9 &       4.0 & 15.9 & 17.3 &  21.5  &  17.4 &  &$-$14.70 &    diff   & O8  V& near F2H7\\
HII 6   & 10:02:52.03 & -26:09:05.3 &       0.8 & 14.2 &    16.6 &  20.8  & 15.5 &     & $-$14.02 &comp tail & O6-O7 V& F2H3 \\   
HII 7   & 10:02:52.46 & -26:09:24.6 &      1.4 & 14.4 &    16.5  & 20.7 & 15.8 &$-$14.11 & $-$14.10&  comp  & O6-O7 V  & near F2H6\\     
HII 8   & 10:02:56.28 & -26:08:58.3 &    17.0 & 15.3 &      &  &  15.7 &    &$-$14.45& ext &   clump& HII1 RMc; Ho12; F1H4,H12 \\
HII 9  &  10:02:58.11 & -26:08:52.5 &    5.0  & 17.2 &   17.2 &  21.4 & 17.2 &    & $-$15.21 &   diff  & O8-O9 V & \\
HII 10 & 10:02:58.91 & -26:08:48.3 &   13.0 &16.1 &   15.9 &  20.1 & 16.1 &    & $-$14.79 & ext & O7 III&  HII2 RMc; Ho11\\
HII 11 & 10:02:59.16 & -26:09:17.5 & $<$0.8 &15.0 &  17.0 & 21.2 & 16.0 & $-$14.32 & $-$14.33 &  unres  & O8-O9  V &    PN2 RMc \\         
HII 12 & 10:02:59.52 & -26:09:46.6 &    4.7  &15.9 &    16.7  & 20.9  & 16.9 &      & $-$14.69 & diff   &O7-O8 V& HII3 RMc\\
HII 13 & 10:03:00.12 & -26:08:53.3 &   20.0 & ---    & ---  & ---   & --- &     &--- & ext loop& clump & Ho8; HII4 RMc; Hu5 \\
HII 14 & 10:03:00.19 & -26:09:28.7 &    0.8 & 15.4 &    16.0 &  20.2  & 15.6  &      &$-$14.48 & comp & O6 III  &     PN3 RMc \\ 
HII 15 & 10:03:01.34 & -26:09:27.2 &   $<$0.8   & 14.3 &  16.5  & 20.7 &  15.8 & $-$14.11 & $-$14.06 & unres  & O5-O6 V  &   PN4 RMc\\ 
HII 16 & 10:03:01.37 & -26:08:38.3 &    8.5 & 15.3 &   15.0   & 19.2 & 15.6 &        & $-$14.45 & ext & O Ia & HII5 RMc, F1H3\\
HII 17 & 10:03:01.97 & -26:08:37.5 &   $<$0.8 & 15.2 &   17.2  & 21.4 &16.1 & $-$14.42 & $-$14.42 &  unres  & O8-O9 V &      PN5 RMc\\
HII 18 & 10:03:03.77 & -26:09:20.7 &   13.0 & 15.4 &   15.3  & 19.5  &15.2 &        & $-$14.48 & ext & clump &HII6 RMc, D2H5 \\
HII 19 & 10:03:04.46 & -26:08:43.9 &    3.0 & 15.7 &   ---  & --- & ---  &  & $-$14.61 &diff  & ---  &     HII7 RMc\\
HII 20 & 10:03:04.54 & -26:09:20.7 &    1.6 & 14.3 &   16.4  & 20.6 & 15.7 & $-$14.09 &$-$14.05 &  comp  & O6-O7 V   &   PN7 RMc\\       
HII 21 & 10:03:05.16 & -26:09:23.8 &    4.0 & 16.0 &   15.5  & 19.7  & 16.1 &        & $-$14.73 &diff & O5 III & HII8 RMc \\
HII 22 & 10:03:08.50 & -26:10:02.5 &  60.0 & ---  &  ---  & ---  & ---  &    & --- & arc  & --- &  D2H4; Hu3\\
HII 23 & 10:03:10:67 & -26:09:25.9 &     8.0 & 18.0 &   16.1  & 20.3  &  16.1 &      &$-$15.54& diff & O4-O5  V&\\
HII 24 & 10:03:11.50 & -26:10:22.4 &   1.2 & 17.2 &   17.4  & 21.6  & 17.5 &      & $-$15.21 & diff  & O8-O9 V  & \\        
HII 25 & 10:03:11.78 & -26:09:47.3 &    2.6 & 15.7 &   16.6  & 20.8 &  16.6 &      & $-$14.61 & diff  & O6-O7 V    &    \\
HII 26 & 10:03:12.64 & -26:08:52.1 &    6.0 & 15.7 &   15.5  & 19.7 &  15.8 &      & $-$14.60 &diff ext & O5 III & HII9 RMc; Ho3; Hu8\\
HII 27 & 10:03:12.84 & -26:09:54.9 &    2.0 & 16.8 &   17.9  & 22.1 & 17.2 &      & $-$15.05 & diff  & O8-O9  V  & \\               
HII 28 & 10:03:13.32 & -26:08:50.0 &    2.4 & 15.6 &   17.0  & 21.2 & 17.1 &      & $-$14.58 & diff  & O8  V  &  \\
HII 29 & 10:03:17.48 & -26:10:00.5 &    6.0 & 16.7 &   16.9  & 21.1 & 17.1 &   & $-$14.99 & diff  & O7-O8 V &  HII10 RMc, Ho2\\
HII 30 & 10:03:17.71 & -26:09:59.4 &    0.8 & 16.0 &   17.5  & 21.7 & 17.6 &     & $-$14.74 & comp  & O8-O9 V& near HII10 RMc\\ 
HII 31 & 10:03:22.00 & -26:08:53.4 &    2.5 & 15.8 &   17.6  & 21.8 & 17.6 &     & $-$14.65 &diff  & O8-O9 V & D1H6\\
HII 32 & 10:03:22.92 & -26:10:27.4 &    1.5 & 15.6 &   16.5  & 20.7 & 16.2 & $-$14.58 & $-$14.55 & comp &  O7 V & \\
HII 33 & 10:03:25.50 & -26:10:54.7 &  18.0 & 17.5 &   16.3  & 20.5 & 16.9 &      & $-$15.33 & ext & O5-O6 V&\\
HII 34 & 10:03:25.80 & -26:09:08.9 &   0.9  & 17.4 &   17.0  & 21.2 & 16.7 & $-$15.23 &  $-$15.29 & comp &  O7 V & \\    
HII 35 & 10:03:27.19 & -26:10:22.4 &  19.0 &17.7  &   17.3  & 21.5 & 17.5 &     & $-$15.39 &ext cmplx  & clump & Ho1, Hu1 \\ 
HII 36 & 10:03:27.90 & -26:09:36.4 &   2.5  & 17.3 &   17.4  & 21.6 & 16.4 &     & $-$15.25 & diff  & O8-O9 V & \\
HII 37 & 10:03:27.96 & -26:09:19.9 &    0.9 & 13.6 &   16.7  & 20.9 & 15.2 &     & $-$13.75 & comp &  O7-O8 V  & \\            
HII 38 & 10:03:28.00 & -26:10:22.8 &    3.0 & 16.9 &   16.8  & 21.0 & 17.6 &     & $-$15.09 & diff  & O7-O8 V&  F5H1\\
HII 39 & 10:03:28.08 & -26:09:14.8 &    3.0 & 17.5 &   17.3  & 21.5 & 17.8 &     & $-$15.33 & diff  & O8-O9 V & F5H2 \\
HII 40 & 10:03:28.51 & -26:09:21.8 &    0.8 & 13.2 &  15.4   & 19.6 & 14.4 & $-$13.55 & $-$13.57 & comp& clump & near F5H4,H3,H5 \\      
\hline
\multicolumn{12}{l}{(a) V stellar magnitude of the ionizing star, as derived relative to DKI photometry.}\\
\multicolumn{13}{l}{(b) Flux in erg cm$^{-2}$ s$^{-1}$, column 9: from spectroscopy; column 10: from expression (1).}\\
\multicolumn{13}{l}{(c) Nomenclature: comp: compact, cmplx: complex, diff: diffuse, ext: extended, unres: unresolved.}\\
\multicolumn{13}{l}{(d) References: RMc, Richer \& McCall 1992; Ho: Hodge 1969; Hu: Hunter et al 1993; F3H1, F5H4, D1H6, ... : Bresolin et al. 1993.}\\
\end{longtable}
\end{landscape}
}

\end{document}